\documentclass[10pt,twocolumn,aps,pra,superscriptaddress,footinbib,floatfix]{revtex4-1} 
\usepackage{amsmath,amssymb,graphicx,times,color,upgreek}
\usepackage[colorlinks, linkcolor=blue, citecolor=blue, urlcolor=blue, breaklinks=true]{hyperref}

\newcommand{\eref}[1]{Eq.~(\ref{eq:#1})}
\newcommand{\erefs}[1]{Eqs.~(\ref{eq:#1})}
\newcommand{\Eref}[1]{Equation~(\ref{eq:#1})}
\newcommand{\Erefs}[1]{Equations~(\ref{eq:#1})}
\newcommand{\fref}[1]{Fig.~\ref{fig:#1}}
\newcommand{\Fref}[1]{Figure~\ref{fig:#1}}

\DeclareMathOperator{\Tr}{Tr}

\begin{document}

\title{Manipulating the flow of thermal noise in quantum devices}

\author{Shabir Barzanjeh}
\email[Electronic address:\ ]{shabir.barzanjeh@ist.ac.at}
\affiliation{Institute of Science and Technology Austria, 3400 Klosterneuburg, Austria}
\author{Matteo Aquilina}
\affiliation{National Aerospace Centre, Luqa LQA\,9023, Malta}
\author{Andr\'e Xuereb}
\email[Electronic address:\ ]{andre.xuereb@um.edu.mt}
\affiliation{Department of Physics, University of Malta, Msida MSD\,2080, Malta}

\begin{abstract}
There has been significant interest recently in using complex quantum systems to create effective non-reciprocal dynamics. Proposals have been put forward for the realization of artificial magnetic fields for photons and phonons; experimental progress is fast making these proposals a reality. Much work has concentrated on the use of such systems for controlling the flow of signals, e.g., to create isolators or directional amplifiers for optical signals. In this paper, we build on this work but move in a different direction. We develop the theory of and discuss a potential realization for the controllable flow of thermal noise in quantum systems. We demonstrate theoretically that the unidirectional flow of thermal noise is possible within quantum cascaded systems. Viewing an optomechanical platform as a cascaded system we here that one can ultimately control the direction of the flow of thermal noise. By appropriately engineering the mechanical resonator, which acts as an artificial reservoir, the flow of thermal noise can be constrained to a desired direction, yielding a thermal rectifier. The proposed quantum thermal noise rectifier could potentially be used to develop devices such as a thermal modulator, a thermal router, and a thermal amplifier for nanoelectronic devices and superconducting circuits.
\end{abstract}
\maketitle

\section{Introduction}
The control of thermal noise in complex systems has straightforward applications to the miniaturization of technology; as devices become smaller and smaller it is essential to steer thermal noise away from hot spots towards sinks where it may be disposed of (see, e.g., Ref.\ \cite{2016NatPh..12..460P}). Recently a significant effort has emerged that is devoted to design a new generation of thermal-based nanoscale devices such as thermal rectifiers~\cite{2002PhRvL..88i4302T, 2007PhRvB..76b0301Y, 2008PhRvL.100j5901S, 2008NJPh...10h3016S, 2011ApPhL..98k3106B, 2014ApOpt..53.3479N, 2014PhRvE..89f2109W}, thermal logic gates~\cite{2007PhRvL..99q7208W}, thermal diodes~\citep{2004PhRvL..93r4301L, 2017PhRvE..95b2128O}, and thermal transistors~\cite{2008JPSJ...77e4402L, 2016PhRvL.116t0601J, Thierschmann2017}. When quantum systems are coupled together, the thermal noise associated with the reduced state of each component is affected by the coupling, leading to a flow a thermal noise~(see Appendix); controlling this thermal noise is essential in the context of quantum technologies, such as quantum computers~\cite{2010Natur.464...45L} and simulators~\cite{2014RvMP...86..153G}, especially because of the fragility of quantum states and quantum correlations~\cite{2012JPhA...45x4002S} that is well-known from the literature. Coupled quantum systems can also be used to transfer signals; a signal input to one quantum system can appear at the output of another~(see Appendix). A basic building block for controlling how such signals flow around a complex system takes the form of devices that are non-reciprocal, in which transmission of a signal from one point to another is qualitatively different~\cite{1969ITMTT..17.1087W, 1983OptL....8..560K, 2004RPPh...67..717P, 2010PhRvA..81c2107L, 2012AnPhy.327.1050D}, in amplitude or phase, from transmission in the reverse direction. An interesting line of research has emerged recently that aims to use complex mechanical, electromagnetic, or quantum-optical systems to create effective optical isolators~\cite{2013NaPho...7..579J} or other kinds of non-reciprocity~\cite{2017Natur.541..473L}. Several theoretical analyses~\cite{2009PhRvL.102u3903M, 2011NatPh...7..311K, 2012OExpr..20.7672H, 2012NaPho...6..782F, 2014NJPh...16j3027R, 2015PhRvX...5b1025M, 2015PhRvA..91e3854X, 2015PhRvX...5c1011P, 2016PhRvA..93b3827X, 2017PhRvA..96a3808T, 2016NJPh...18k3029W, PhysRevApplied.7.064014, 2017OExpr..2518907L} of such systems have been published and experimental demonstrations~\cite{2011Sci...333..729F, 2012Sci...335Q..38F, 2012Sci...335R..38F, 2011NaPho...5..549K, 2011NaPho...5..758B, 2012PhRvL.109c3901L, 2013OptL...38.1259F, 2013PhRvX...3c1001A, 2014NaPho...8..701T, 2014NatPh..10..923E, 2016NaPho..10..657S, 2016NatCo...713662R, Bernier2017, 2016Sci...354.1577S, 2017arXiv170102699X, 2017NatPh..13..465F, 2017Natur.542..461C, 2017PhRvP...7b4028L, 2017PhRvX...7c1001P, 2017PhRvA..95e3822M, Barzanjeh2017} reported, illustrating a rich variety of mechanisms for achieving the desired non-reciprocity. In their simplest form, several such mechanisms are based on coupled quantum systems that also share a common bath~\cite{2015PhRvX...5b1025M}. These can be conceptually connected to techniques discussed several years ago under the guise of cascaded quantum systems~\cite{Gardiner2000}.

\begin{figure}[b]
 \includegraphics[width=0.9\linewidth]{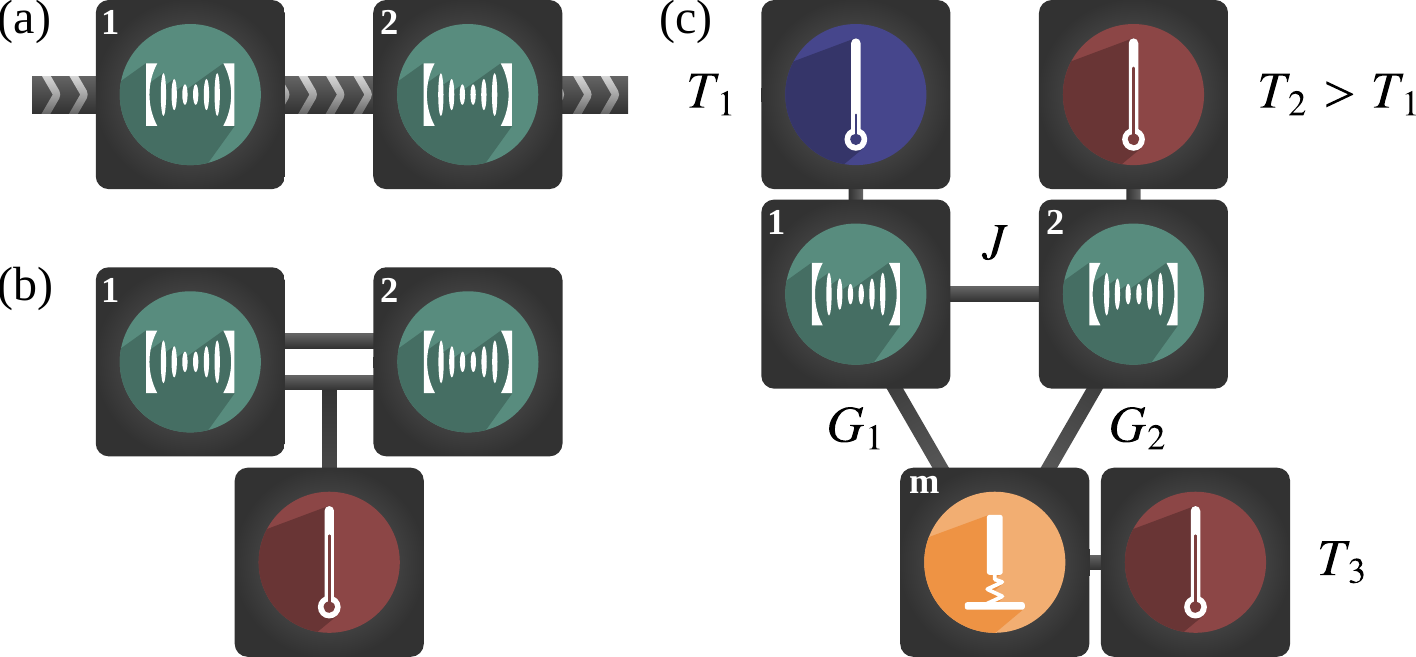}
 \caption{In our quantum optics model (a)~two harmonic oscillators, e.g., electromagnetic (optical or microwave) cavities, are arranged such that the output from system 1 is the input of system 2. (b)~Equivalently, the two systems are connected via a coherent hopping term and share a common heat bath. (c)~A physical realization of a thermal rectifier; a mechanical system is coupled to two electromagnetic cavities and a heat bath, is proposed as a realization. In this model, the two systems are also connected to their own heat baths.}
 \label{fig:Model}
\end{figure}

In this paper we will combine cascaded quantum systems, non-reciprocal devices, and controlling the flow of thermal noise to achieve a thermal rectifier. We analyze a quantum system consisting of two fields between which we set up non-reciprocal transport. Our analysis differs from what is known in the literature because we are interested not in the transport of signals, but in the transport of thermal noise between the two fields. We also use recently-developed techniques~\cite{2015PhRvA..92a3844P, 2016NJPh...18a3009P, 2016JSMTE..06.3203P} for analysing the flow of excitations between quantum systems and their heat baths to better understand how our system manipulates the flow of thermal noise. Our work thus considers thermal noise not as a nuisance complicating our analysis, but as the object of that analysis. Our results show that the temperature of a third bath can be used to \emph{increase or decrease} the thermal noise of one system without affecting the other, paving the way to quantum thermal transistors.

We proceed as follows. First, we describe an effective quantum optics model based on the cascaded quantum systems formalism [\fref{Model}(a)]. This yields general expressions that have a transparent physical meaning. We then develop an optomechanical model where a mechanical oscillator is coupled to two electromagnetic cavities [\fref{Model}(b,c)]. We can show that these two systems behave identically with respect to quantum states that are broadband compared to coherent signals but still contained within the bandwidth of the mechanical oscillator. This allows us to use our general expressions to derive conclusions about this specific system. We then discuss how the effect we explore manifests itself in experiment, with reference to achievable parameters. Finally, we conclude with a short discussion on the implications of our model.

\section{Effective quantum optics model}
To start off, we briefly summarize what is known about cascaded quantum systems; our aim is to build a model for the system shown in \fref{Model}(a) to discuss its operation as a non-reciprocal thermal device. We start by considering two harmonic oscillators, associated with annihilation operators $\hat{c}_1$ and $\hat{c}_2$, respectively. Let $\hat{H}_\text{sys}$ govern the free dynamics of these systems. We \emph{impose} cascaded dynamics (see Sec.\ 12.1 of Ref.\ ~\cite{Gardiner2000}) onto these systems, i.e., we assume that the output of oscillator $1$ forms the input of oscillator $2$ via some channel, whereas the output of oscillator $2$ does not feed back into oscillator $1$. Define $\gamma_1>0$ and $\gamma_2>0$ as the coupling rates of the two systems, respectively, to this channel. Together with the two standard input--output relations $\hat{c}_{\text{out},i}=\hat{c}_{\text{in},i}+\sqrt{\gamma_i}\hat{c}_i$, where $i=1,2$, we must therefore add the restriction $\hat{c}_{\text{in},2}=\hat{c}_{\text{out},1}$. Next we can follow Ref.\ \cite{Gardiner2000} in obtaining the Langevin equations governing the dynamics of this system, converting them into It\={o} stochastic differential equations, and from there deriving a master equation. In the following we denote by $\bar{N}_3$ the average occupation number of an effective common bath, and we take there to be no classical driving field associated with this bath. This master equation can be rewritten in Lindblad form to yield $\dot{\rho}=-\tfrac{\imath}{\hbar}[\hat{H}_\text{sys}+\hat{H}_\text{hop},\rho]+(\bar{N}_3+1)\kappa_3\mathcal{D}_{\hat{c}_3}[\rho]+\bar{N}_3\kappa_3\mathcal{D}_{\hat{c}_3^\dagger}[\rho]$, where $\mathcal{D}_{\hat{c}}[\rho]=\hat{c}\rho\hat{c}^\dagger-\tfrac{1}{2}\{\rho,\hat{c}^\dagger\hat{c}\}$ is the standard dissipative Lindblad term, $\hat{H}_\text{hop}=\tfrac{\imath\hbar}{2}\sqrt{\gamma_1\gamma_2}(\hat{c}_1^\dagger\hat{c}_2-\hat{c}_1\hat{c}_2^\dagger)$ is a hopping Hamiltonian, $\kappa_3=\gamma_1+\gamma_2$ is a collective damping rate, and $\hat{c}_3=(\sqrt{\gamma_1}\hat{c}_1+\sqrt{\gamma_2}\hat{c}_2)/\sqrt{\kappa_3}$ is a collective bosonic annihilation operator that obeys $[\hat{c}_3,\hat{c}_3^\dagger]=1$. The physical content of this master equation is rather straightforward: To produce the non-reciprocal effect required of a cascaded system, the two oscillators must be coupled by a direct coherent hopping term as well as to a common bath; see \fref{Model}(b). To account for an arbitrary phase $\phi$ in the hopping between the two oscillators, we replace $\hat{c}_2\to e^{\imath\phi}\hat{c}_2$ throughout, yielding $\hat{H}_\text{hop}=\tfrac{\imath\hbar}{2}\sqrt{\gamma_1\gamma_2}(e^{\imath\phi}\hat{c}_1^\dagger\hat{c}_2-e^{-\imath\phi}\hat{c}_1\hat{c}_2^\dagger)$ and $\hat{c}_3=(\sqrt{\gamma_1}\hat{c}_1+\sqrt{\gamma_2}e^{\imath\phi}\hat{c}_2)/\sqrt{\kappa_3}$. This master equation results in equations of motion that are maximally non-reciprocal with respect to $\hat{c}_1$ and $\hat{c}_2$, which is due to a coherent cancellation (addition) of the hopping between the direct term and through the bath in the direction $2\to1$ ($1\to2$). The phase-matching condition required to ensure this cancellation or addition is encoded in a $-$ sign in the coherent hopping Hamiltonian, compared to a $+$ sign in the dissipation-related operator $\hat{c}_3$. For further generality, we must add terms to this master equation. First, we modify the hopping Hamiltonian to $\hat{H}_\text{hop}=\tfrac{\imath\hbar}{2}\sqrt{\gamma_1\gamma_2}(e^{\imath\phi}\hat{c}_1^\dagger\hat{c}_2-e^{-\imath\phi}\hat{c}_1\hat{c}_2^\dagger)+\hbar(F\hat{c}_1^\dagger\hat{c}_2+F^\ast\hat{c}_1\hat{c}_2^\dagger)$, where $F$ is an arbitrary complex constant; full non-reciprocity requires $F=0$. Second, we add a bath for each oscillator:
\begin{multline}
\dot{\rho}=-\tfrac{\imath}{\hbar}[\hat{H}_\text{sys}+\hat{H}_\text{hop},\rho]\\
+\textstyle\sum_{i=1}^3\bigl\{(\bar{N}_i+1)\kappa_i\mathcal{D}_{\hat{c}_i}[\rho]+\bar{N}_i\kappa_i\mathcal{D}_{\hat{c}_i^\dagger}[\rho]\bigl\}.
\label{eq:FullME}
\end{multline}
In the following we will use this master equation to describe any system composed of two oscillators that are coupled directly to one another, to a common thermal bath, and to two individual thermal baths [\fref{Model}(c)]. We will show that an effective model where the coupling between two electromagnetic cavities and their common bath are induced by a third, mechanical, mode is equivalent to the one described here.

To proceed, we convert the master equation to its equivalent quantum Langevin equations~(see Appendix): We derive the mean-field equations of motion from \eref{FullME}, obtain the operator equations by adding noise terms using the fluctuation--dissipation theorem, and then Fourier-transform to the frequency domain:
\begin{multline}
-\imath\omega\begin{pmatrix}
\hat{c}_1 \\
\hat{c}_2
\end{pmatrix} = \begin{bmatrix}
-\imath\omega_1-\tfrac{\gamma_1+\kappa_1}{2} & -\imath F \\
-\imath F^\ast-\sqrt{\gamma_1\gamma_2}e^{\imath\phi} & -\imath\omega_2-\tfrac{\gamma_2+\kappa_2}{2}
\end{bmatrix}\begin{pmatrix}
\hat{c}_1 \\
\hat{c}_2
\end{pmatrix}\\
+\begin{pmatrix}
\sqrt{\kappa_1}\hat{c}_{\text{in},1} \\
\sqrt{\kappa_2}\hat{c}_{\text{in},2}
\end{pmatrix}+\begin{pmatrix}
\sqrt{\gamma_1} \\
\sqrt{\gamma_2}e^{\imath\phi} \\
\end{pmatrix}\hat{c}_\text{in,3}.
\label{eq:Langevin}
\end{multline}
Under the white-noise assumption, these zero-mean noise operators are such that $\langle\hat{c}_{\text{in},i}(t)\hat{c}_{\text{in},j}^\dagger(t^\prime)\rangle=(\bar{N}_i+1)\delta_{i,j}\delta(t-t^\prime)$, $\langle\hat{c}_{\text{in},i}^\dagger(t)\hat{c}_{\text{in},j}(t^\prime)\rangle=\bar{N}_i\delta_{i,j}\delta(t-t^\prime)$, and $\langle\hat{c}_{\text{in},i}(t)\hat{c}_{\text{in},j}(t^\prime)\rangle=0$ ($i,j=1,2,3$). Since \eref{Langevin} is a linear system of equations, a full description of the state at any point in time requires only the first and second moments of the quadrature operators $\hat{x}_i=(\hat{c}_i+\hat{c}_i^\dagger)/\sqrt{2}$ and $\hat{p}_i=-\imath(\hat{c}_i-\hat{c}_i^\dagger)/\sqrt{2}$ ($i=1,2$). It can be shown that the covariance matrix $V$ of this system obeys the Lyapunov equation $\dot{V}=A\cdot V+V\cdot A^\text{T}+N$, where the drift matrix $A$ is related to the matrix in the first term of \eref{Langevin} and the noise matrix $N$ is related to the second and third terms of this same equation. When the eigenvalues of $A$ all have negative real parts, a unique solution to $V$ exists. In our case, we define $\bar{n}_i=\langle\hat{c}^\dagger_i\hat{c}_i\rangle$ and $\Delta:=\omega_2-\omega_1$, and simplify our expressions by taking $\kappa_1=\kappa_2=\gamma_1=\gamma_2=:\kappa$. We want to compare our system to one in which the two oscillators lack any direct coupling or common bath. Simply removing the common bath and the link between the oscillators fundamentally alters the nature of the system, as it changes the number of baths each oscillator is connected to. For a physically meaningful comparison we must modify the bath parameters appropriately. In this \emph{disconnected} scenario, which is physically equivalent to taking $\lvert\Delta\rvert\to\infty$ in the above expressions whilst keeping $F$, $\kappa$, and $\bar{N}_i$ ($i=1,2,3$) fixed, the steady-state occupation numbers are $\bar{m}_i=\tfrac{1}{2}(\bar{N}_i+\bar{N}_3)$ ($i=1,2,3$); note that the $\bar{m}_i$ are independent of $F$ and that $\bar{m}_3=\bar{N}_3$. Define $\Delta n_i:=\bar{n}_i-\bar{m}_i$ ($i=1,2$) to quantify the difference between the two scenarios, whose explicit expressions we reproduce elsewhere~(see Appendix). For simplicity let us look at the maximally non-reciprocal case ($F=0$), whereby
\begin{figure}[t]
 \includegraphics[width=\linewidth]{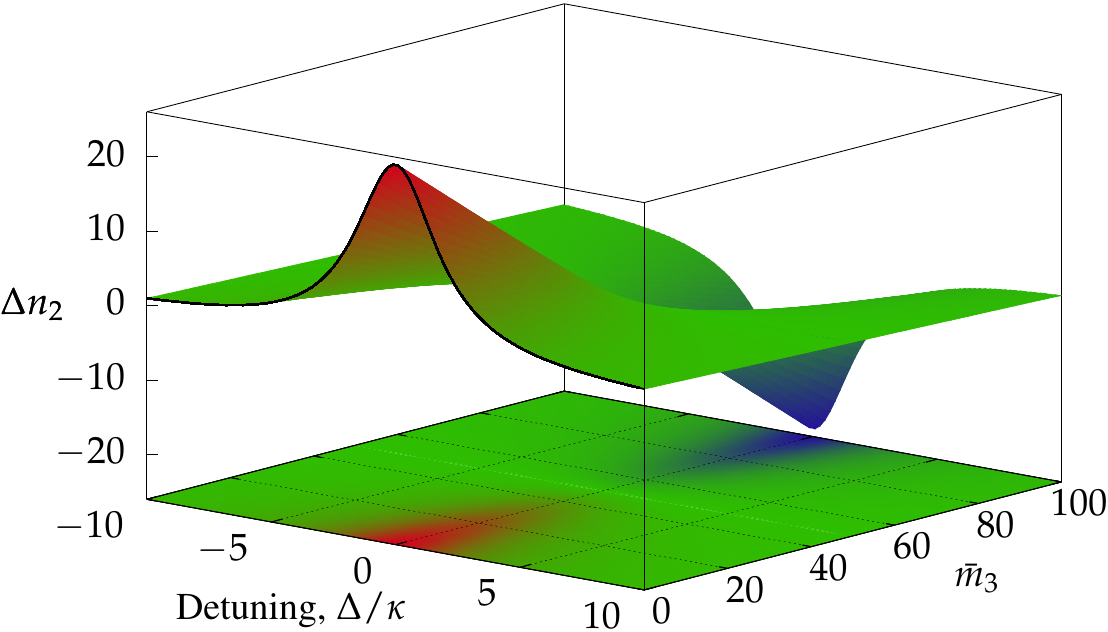}
 \caption{(Color online) Change in occupation number of the second oscillator, $\Delta n_2$, as a function of the detuning $\Delta$ between the two oscillators and the occupation number $\bar{m}_3$ of the common bath. Red (blue) regions correspond to increased (decreased) thermal noise. Note that $\Delta n_1=0$ throughout. ($\phi=0$, $\bar{m}_1=50$, $\bar{m}_2=100$.)}
 \label{fig:Dn2}
\end{figure}
\begin{equation}
\Delta n_1=0\ \text{and}\ \Delta n_2=\tfrac{2\kappa^2}{4\kappa^2+\Delta^2}(\bar{m}_1-\bar{m}_3).
\end{equation}
This very clearly shows that, whatever the value of $\bar{m}_1-\bar{m}_2$, we find an \emph{increase} (\emph{decrease}) in $\bar{n}_2$ over the disconnected case for $\bar{m}_1>\bar{m}_3$ ($\bar{m}_1<\bar{m}_3$), whereas $\bar{n}_1$ is unaffected by the presence of the other oscillator. It is interesting to note that this conclusion remains unchanged if we have $\bar{m}_2=\bar{m}_1$. In other words, even if the two oscillators equilibrate to the same temperature in the disconnected case, the channel will cause an excess or depleted the flow of thermal noise towards oscillator $2$ that depends only on the temperature difference between oscillators $1$ and $3$. \Fref{Dn2} shows that the temperature of the common bath can be used as a control knob to modulate the flow of thermal noise into or out of the second oscillator. Note that, for $F=0$ the temperature of the first oscillator is unaltered. The temperature of the second oscillator can be lower (blue), the same (green), or higher (red) in comparison to the disconnected scenario depending on the temperature of the common bath and $\Delta$ which, e.g., can be chosen to reduce the flow of thermal noise into oscillator 2 even when all coherent signals flow from oscillator 1 to 2. The case for $F\neq0$ is shown in Fig.~S.1 of the Appendix. Regardless of the temperature difference between the two oscillators \emph{and the direction of signal flow}, the thermal noise flowing into the second oscillator can be increased or decreased.

We next turn our attention to an experimentally-feasible optomechanical platform that can realize this model. We shall use terminology related to platforms operating in optical domain, but all our results hold identically for microwave-based systems. Our results are important for interfacing with such systems, since the thermal occupation of the electromagnetic field at microwave frequencies is often non-negligible.

\section{Optomechanical realization}
Our aim in this section is to employ a mechanical degree of freedom interacting with two optical fields, acting as a controllable reservoir. The result is an optomechanical system that works as a thermal rectifier, with the temperature of the mechanical oscillator bath controlling the steady-state temperature of the second optical field. A schematic realization of this optomechanical system is sketched in \fref{Model}(c). Here we consider an optomechanical platform consisting of two optical cavities with resonance frequencies $\omega_i$ ($i=1,2$), which interact simultaneously with a mechanical resonator with frequency $\omega_\text{m}$, and where the single-photon optomechanical coupling strength between the oscillator and the $i$\textsuperscript{th} cavity is $g_i$ ($i=1,2$). The direct photon hopping rate between the cavities is denoted by $J$, which is assumed real for simplicity. The Hamiltonian governing the unitary evolution of this system is given by~\cite{PhysRevA.84.042342, PhysRevLett.109.130503, Barzanjeh2017}
\begin{multline}
\hat{H}=\hbar\omega_\text{m}\hat{b}^\dagger\hat{b}+\textstyle\sum_{i=1,2}\hbar\bigl[\omega_i \hat{a}^\dagger_i\hat{a}_i+g_i\bigl(\hat{b}+\hat{b}^\dagger\bigr)\hat{a}^\dagger_i\hat{a}_i\bigr]\\
+\hbar J\bigl(\hat{a}_1\hat{a}_2^\dagger+\hat{a}_1^\dagger\hat{a}_2\bigr)+\textstyle\sum_{i=1,2}\hbar\mathcal{E}_i\big(\hat{a}_ie^{-\imath\omega_\text{d}t}+\text{h.c.}\bigr),
\label{eq:OPMHamiltonian}
\end{multline}
where $\hat{a}_i$ (with $\bigl[\hat a_i,\hat a_j^{\dagger}\bigr]=\delta_{ij}$) are the annihilation operators of the cavity fields and $\hat{b}$ is the mechanical annihilation operator. The first and second terms of \eref{OPMHamiltonian} describe the free Hamiltonians of the mechanical and cavity fields, respectively; the third term indicates the optomechanical coupling between the cavities and the mechanical resonator; and the fourth term shows the cavity--cavity photon hopping. The last term represents the driving of each cavity $i$ by a coherent electromagnetic field with frequency $\omega_\text{d}$, which we assume to be the same for both cavities, and amplitude $\mathcal{E}_i$. We note that our analysis also applies to systems where two mechanical modes are used to generate non-reciprocal coupling between two electromagnetic cavities. Recent realizations of such systems~\cite{Bernier2017, 2017NatPh..13..465F, 2017PhRvX...7c1001P, Barzanjeh2017} illustrate the feasibility of implementing non-reciprocal transport of thermal noise and signals.

In a rotating frame with respect to $\omega_\text{d}$, and after adding losses by means of dissipative Lindblad terms as in the preceding section, the dynamics of the system can be fully characterized by the quantum Langevin equations of motion ($i=1,2$)
\begin{subequations}
\begin{align}
\dot{\hat{a}}_i&=-\bigl(\imath\Delta_i+\tfrac{\kappa_i}{2}\bigr)\hat{a}_i-\imath J\hat{a}_{\bar i}-\imath g_i\hat{a}_i\bigl(\hat{b}+\hat{b}^\dagger)+\mathcal{E}_i\nonumber\\
 &\qquad+\sqrt{\kappa_i}\hat{a}_{\text{in},i},\\
\dot{\hat{b}}&=-\bigl(\imath\omega_\text{m}+\tfrac{\gamma_\text{m}}{2}\bigr)\hat{b}-\imath\textstyle\sum_{i=1,2}g_i\hat{a}_i^\dagger\hat{a}_i+\sqrt{\gamma_\text{m}}b_{\text{in},\text{m}},
\end{align}
\label{eq:QLE}%
\end{subequations}
where $\Delta_i:=\omega_i-\omega_\text{d}$, $\bar{1}=2$, and $\bar{2}=1$. Here, $\kappa_i:=\kappa_{\text{int},i}+\kappa_{\text{ext},i}$ are the linewidths of the cavities in which $\kappa_{\text{int},i}$ and $\kappa_{\text{ext},i}$ are the intrinsic and extrinsic linewidths, respectively. Intrinsic losses and input quantum noise are associated with the zero-mean noise operators $\hat{a}_{\text{int},i}$ and $\hat{a}_{\text{ext},i}$, respectively; we can conveniently define $\hat{a}_{\text{in},i}:=\bigl(\sqrt{\kappa_{\text{ext},i}}\hat{a}_{\text{ext},i}+\sqrt{\kappa_{\text{int},i}}\hat{a}_{\text{int},i}\bigr)/\sqrt{\kappa_i}$. The damping of the mechanical resonator is given by $\gamma_\text{m}$. The zero-mean quantum fluctuations $\hat{a}_{\text{in},i} $ and $\hat{b}_{\text{in},\text{m}}$ satisfy the usual white noise correlations~(see Appendix). \Erefs{QLE} can be solved linearization around the classical steady state of the system. We define the zero-mean cavity field fluctuation operators $\delta\hat{a}_i:=\hat{a}_i-\alpha_i$ where $\alpha_i=2\mathcal{E}_ie^{\imath\phi_i}/\sqrt{4\Delta_i^2+\kappa_i^2}$ are the steady-state solutions, ignoring a small change in $\Delta_i$ due to a static shift in the position of the mechanical oscillator, and assuming $\lvert\alpha_i\rvert\gg1$. 

If the driving frequencies are chosen such that $\Delta_i\approx\omega_\text{m}$ and the system is in the sideband-resolved regime, i.e., $\omega_\text{m}\gg\kappa_i$, it is possible to use the rotating-wave approximation to drop the rapidly-rotating terms oscillating at $\pm\omega_\text{m}$. This allows to eliminate the mechanical degree of freedom, whereby the equations can be approximated in the frequency domain by
\begin{widetext}
\begin{equation}
-\imath\omega\begin{pmatrix}
  \delta\hat{a}_1 \\
  \delta\hat{a}_2
 \end{pmatrix}=\begin{bmatrix}
  -\imath\Delta_1-\frac{\kappa_1}{2}-G_1^2\chi_\text{m}(\omega) & -\imath J-\chi_\text{m}(\omega)G_1G_2e^{-\imath\phi} \\
  -\imath J-\chi_\text{m}(\omega)G_1G_2e^{\imath\phi}&-\imath\Delta_2-\frac{\kappa_2}{2}-G_2^2\chi_\text{m}(\omega)
 \end{bmatrix}\begin{pmatrix}
  \delta\hat{a}_1 \\
  \delta\hat{a}_2
 \end{pmatrix}+\begin{pmatrix}
  \sqrt{\kappa_1}\hat{a}_{\text{in},1} \\
  \sqrt{\kappa_2}\hat{a}_{\text{in},2}
 \end{pmatrix}+\begin{pmatrix}
  G_1\sqrt{\gamma_\text{m}}\tilde{\chi}_\text{m}(\omega) \\
  G_2\sqrt{\gamma_\text{m}}\tilde{\chi}_\text{m}(\omega)e^{\imath\phi}
 \end{pmatrix}\hat{b}_{\text{in},\text{m}}
 \label{eq:OMfreqdom}
\end{equation}
\end{widetext}
where $G_i=g_i\alpha_i$ is the effective optomechanical coupling rate and the mechanical susceptibility is defined as $\chi_\text{m}(\omega)=1\big/\bigl[\gamma_\text{m}/2-\imath(\omega-\omega_\text{m})\bigr]$. To simplify matters, we chose the phase reference such that $G_1$ is real and set $G_2\to G_2e^{\imath\phi}$ (where the $G_2$ on the right-hand side is real). We also defined $\tilde{\chi}_\text{m}(\omega):=\chi_\text{m}(\omega)\lvert\chi_\text{m}(\Omega)\rvert/\chi_\text{m}(\Omega)$, where $\Omega$ is some frequency of interest. This procedure detailed elsewhere~(see Appendix).

\Eref{OMfreqdom} reveals that in general the photon hopping between cavities is not symmetric---note that the off-diagonal terms of the drift matrix on the right-hand side of the equation are \emph{not} complex conjugates of one another. This means that by properly choosing the system parameters one can break the reversibility of the thermal photon hopping between the cavities and set up a preferred direction for the flow of thermal noise. For example, a situation of full non-reciprocity at frequency $\Omega$, where the photon hopping is entirely suppressed in the direction $2\to1$, may be obtained by choosing the parameters such that $J=\imath\chi_\text{m}(\Omega)G_1G_2e^{-\imath\phi}$.

Consider, now, a quantum state centered around frequency $\Omega$ in the rotating frame and whose bandwidth $\Gamma$ is much smaller than $\gamma_\text{m}$, such that $\chi_\text{m}(\omega)=\tilde{\chi}_\text{m}(\omega)\approx\chi_\text{m}(\Omega)$, constant over the bandwidth of the signal. Under these `large bandwidth' conditions, when $\gamma_\text{m}\gg\Gamma$, all the parameters entering \eref{OMfreqdom} can be held constant, and this equation therefore becomes identical to \eref{Langevin}, with the following replacements: $\omega_i\to\Delta_i+G_i^2\Im\{\chi_\text{m}(\Omega)\}$, $\gamma_i\to2G_i^2\Re\{\chi_\text{m}(\Omega)\}$, and $F\to J-\imath\chi_\text{m}(\Omega)G_1G_2e^{-\imath\phi}$. For example, perfect non-reciprocity requires~$J=G_1G_2\bigl[\bigl(\tfrac{\gamma_\text{m}}{2}\bigr)^2+\bigl(\Omega-\omega_\text{m}\bigr)^2\bigr]^{-1/2}$, with $\phi$ chosen such that $F=0$. A detailed discussion of the equivalence between the two systems is presented elsewhere~(see Appendix). We can therefore apply the formalism developed previously to conclude that any thermal noise in the signal will be suppressed \emph{in one direction} only. By manipulating the properties of the mechanical oscillator, e.g., using an auxiliary optical field, one may control the flow of thermal energy in the electromagnetic signal transmitted between the two cavities. An in-depth analysis~(see Appendix) may be performed to derive the flow of excitations between the system and the three baths it is connected to. Figure~S.2 in the Appendix shows that changing the temperature of \emph{either} resonator does not affect the flow of excitations between the other resonator and its own bath. Any excess flow between the resonators is therefore borne exclusively by their common bath and the link between them. The net flow, given by the sum of the flows to all baths, is shown to be equal to zero, as required for physical consistency.

This proposed thermal rectifier can be implemented using an on-chip microwave electromechanical system based on a lumped-element superconducting circuit with a drumhead capacitor~\cite{Bernier2017, 2017PhRvX...7c1001P} or a dielectric nanostring mechanical resonator~\cite{Barzanjeh2017}. We assume the following experimentally feasible parameters: Optomechanical coupling rates of $G_1=G_2=2\pi\times7$\,kHz, cavities resonant at $2\pi\times5$\,GHz and having damping rates of $\kappa_1=\kappa_2=2\pi\times2$\,MHz, mechanical resonance frequency $\omega_\text{m}=2\pi\times 6$\,MHz and damping rate $\gamma_\text{m}=2\pi\times100$\,Hz. Inductive or capacitive coupling between microwave resonators can yield $J=2\pi\times1$\,MHz. An auxiliary cavity can be used to change the isolation bandwidth $\gamma_\text{m}$. The ambient temperature of the microwave and mechanical resonators can be kept below $10$\,mK by using cryogen-free dilution refrigerators. Optomechanical cooling can be used to cool the mechanical resonator down to ca.\ $0.5$~phonons ($260$\,$\upmu$K). For these parameters the temperature of resonator $2$ is lower with respect to the disconnected case, and depends linearly on that of resonator $1$. Furthermore, the temperature of resonator $1$ is independent of that of resonator $2$.

\section{Conclusions}
We have investigated a generic framework to describe non-reciprocal transport in compound quantum systems. In contrast to several previous studies, we chose to concentrate on the transport of thermal states rather than coherent signals. Our framework can easily be mapped to a prototypical optomechanical realization, which we discussed explicitly in the text. We have also shown how, with parameters typical of present-day microwave optomechanical experiments, the effects we describe should be visible in a proof-of-concept experiment. In the context of quantum measurements and emerging quantum technologies, these techniques and ideas will find use in the manipulation of flow of thermal noise inside quantum devices for phonon-based signal processing and computation, as well as in the construction of quantum-limited amplification systems that perform measurements on sensitive quantum devices without adding thermal noise. Our system can be realized with state-of-the-art technology both in optical~\cite{2017NatPh..13..465F} and microwave~\cite{Barzanjeh2017} domains, and is potentially suited to control the flow of thermal noise in nanoscale devices and to design a new generation of thermal rectifiers, thermal diodes, and transistors. Our work could facilitate noise control and remote cooling of nanoelectronic devices and superconducting circuits using \emph{in situ}-engineerable thermal sinks with possible applications in emerging quantum technologies such as quantum computers and simulators.

\section{Acknowledgments}
We acknowledge funding from the European Union's Horizon 2020 research and innovation program under grant agreement No.\ 732894 (FETPRO HOT). SB acknowledges support under the Marie Sk\l{}odowska-Curie Actions programme, grant agreement No.\ 707438 (MSCA-IF-EF-ST SUPEREOM).

\appendix\onecolumngrid\newpage
\section{Defining temperature, flow of thermal noise, and signal flow}
One can define a temperature of a harmonic oscillator as the temperature of the equivalent thermal state. More precisely, what we call ``temperature'' is the temperature of the heat bath with which a harmonic oscillator reaches equilibrium, such that its equilibrium state is a thermal state with a given mean occupation number. Symbolically, we define the thermal state at temperature $T$ for a harmonic oscillator at frequency $\omega$ through the relation
\begin{equation}
\rho(T):=\frac{e^{-\hbar\omega\hat{a}^\dagger\hat{a}/(k_\text{B}T)}}{\Tr\{e^{-\hbar\omega\hat{a}^\dagger\hat{a}/(k_\text{B}T)}\}},
\end{equation}
where $\hat{a}$ is the annihilation operator for the oscillator and $k_\text{B}$ is Boltzmann's constant. Such a state can uniquely be parametrised either through the temperature $T$ or through its mean occupation number
\begin{equation}
\bar{n}:=\Tr\{\hat{a}^\dagger\hat{a}\,\rho(T)\}=\frac{1}{e^{\hbar\omega/(k_\text{B}T)}-1}.
\end{equation}
Since this relation is one-to-one there is no ambiguity in using either quantity. When the state $\rho$ of a harmonic oscillator is such that $\rho=\rho(T)$ for some $T\geq0$ we refer to this value of $T$ as the temperature of the oscillator. We restrict our discussion to $T\geq0$, which implies that $\bar{n}\geq0$. The mean occupation number of a thermal state is associated with the variance of its quadratures $\hat{x}=(\hat{a}+\hat{a}^\dagger)/\sqrt{2}$ and $\hat{p}=(\hat{a}-\hat{a}^\dagger)/(\imath\sqrt{2})$:
\begin{equation}
\sqrt{\langle\hat{x}^2\rangle-\langle\hat{x}\rangle^2}=\sqrt{\langle\hat{p}^2\rangle-\langle\hat{p}\rangle^2}=\bar{n}+\tfrac{1}{2}.
\end{equation}
This relation between the variances and the mean occupation number allows us to use $\bar{n}$ as a proxy for the thermal noise in the state of the oscillator. When the oscillator is coupled to an output channel, this thermal noise can be observed as noise in the output signal, with amplitude that increases monotonically with $\bar{n}$.

Given two coupled oscillators, each connected to their own heat bath, the flow of thermal noise can be defined qualitatively through its effect on the average occupation number of the reduced thermal states of the two oscillators. In this simple model, the average occupation number of the thermal state describing the oscillator with the cooler bath increases when the two oscillators are coupled; this can be described as a flow of thermal noise to this oscillator. This definition, which has a straightforward physical interpretation, forms the basis of our work.

One can drive either of the oscillators using a monochromatic force, which can be called a coherent ``signal,'' and monitor how that drive affects the state of the other oscillator. In the language of quantum optics, this is equivalent to displacing the state of one oscillator and seeing how that translates to a displacement of the other oscillator. To give this model a more physically-relevant foundation, let us introduce one input--output channels associated with each oscillator; our heat baths can also serve this purpose. One can derive the relevant input--output relations that quantify how a coherent signal in an input channel connected to one oscillator is transferred to an output channel connected to the other oscillator. We refer to this process as signal flow.

\section{Theoretical background of optomechanical systems}
In this section, we explore the theoretical model describing the optomechanical system discussed in the main text. A schematic of the optomechanical thermal noise rectifier is sketched in Fig.~1(c) of the main text, where we consider an optomechanical cavity platform in the form of two cavities, with resonance frequencies $\omega_i$ ($i=1,2$), which interact simultaneously with a mechanical resonator with frequency $\omega_\text{m}$ in which the single-photon optomechanical coupling rate for the interaction between cavity $i$ and the mechanical element is given by $g_i$. The direct photon hopping rate between the cavities is defined by $J$. The Hamiltonian of the system is given by 
\begin{equation}
\hat{H}=\hbar\omega_\text{m}\hat{b}^\dagger\hat{b}+\sum_{i=1,2}\hbar\bigl[\omega_i \hat{a}^\dagger_i\hat{a}_i+g_i\bigl(\hat{b}+\hat{b}^\dagger\bigr)\hat{a}^\dagger_i\hat{a}_i\bigr]
+\hbar J\bigl(\hat{a}_1\hat{a}_2^\dagger+\hat{a}_1^\dagger\hat{a}_2\bigr)+\sum_{i=1,2}\hbar\mathcal{E}_i\big(\hat{a}_ie^{-\imath\omega_{\text{d},i}t}+\text{h.c.}\bigr),
\label{eq:OPMHamiltonianStart}
\end{equation}
where $\hat{a}_i$ (with $[\hat{a}_i,\hat{a}_j^\dagger]=\delta_{i,j}$) is the annihilation operator of cavity $i=1,2$, and $\hat{b}$ is the mechanical annihilation operator. The first and second terms of the Hamiltonian~(\ref{eq:OPMHamiltonianStart}) show the free energy of the mechanical and cavity fields, respectively, while the third term indicates the the optomechanical coupling between the cavities and mechanical resonator. The fourth term stands for the direct cavity--cavity interaction Hamiltonian, where photon hopping occurs with rate $J$. The last term shows that each cavity is driven by a coherent external source with amplitude $\mathcal{E}_i$ and frequency $\omega_{\text{d},i}$. In the rotating frame with respect to the drive frequencies, the above Hamiltonian reduces to
\begin{equation}
\hat{H}=\hbar\omega_\text{m}\hat{b}^\dagger\hat{b}+\sum_{i=1,2}\hbar\bigl[\Delta_i \hat{a}^\dagger_i\hat{a}_i+g_i\bigl(\hat{b}+\hat{b}^\dagger\bigr)\hat{a}^\dagger_i\hat{a}_i\bigr]
+\hbar J\bigl(\hat{a}_1\hat{a}_2^\dagger+\hat{a}_1^\dagger\hat{a}_2\bigr)+\sum_{i=1,2}\hbar\mathcal{E}_i\big(\hat{a}_i+\text{h.c.}\bigr),
\end{equation}
where $\Delta_i:=\omega_i-\omega_{\text{d},i}$, and for simplicity we have assumed $\omega_{\text{d},1}=\omega_{\text{d},2}$.

After adding losses by means of dissipative Lindblad terms, as in the preceding section, the dynamics of the system can be fully characterized by the quantum Langevin equations of motion~\cite{PhysRevA.84.042342, PhysRevLett.109.130503, PhysRevLett.114.080503}
\begin{subequations}
\begin{align}
\dot{\hat{a}}_1&=-\bigl(\imath\Delta_1+\tfrac{\kappa_1}{2}\bigr)\hat{a}_1-\imath J\hat{a}_2-\imath g_1\hat{a}_1\bigl(\hat{b}+\hat{b}^\dagger)+\mathcal{E}_1+\sqrt{\kappa_1}\hat{a}_{\text{in},1},\\
\dot{\hat{a}}_2&=-\bigl(\imath\Delta_2+\tfrac{\kappa_2}{2}\bigr)\hat{a}_2-\imath J\hat{a}_1-\imath g_2\hat{a}_2\bigl(\hat{b}+\hat{b}^\dagger)+\mathcal{E}_2+\sqrt{\kappa_2}\hat{a}_{\text{in},2},\ \text{and}\\
\dot{\hat{b}}&=-\bigl(\imath\omega_\text{m}+\tfrac{\gamma_\text{m}}{2}\bigr)\hat{b}-\imath\sum_{i=1,2}g_i\hat{a}_i^\dagger\hat{a}_i+\sqrt{\gamma_\text{m}}b_{\text{in},\text{m}},
\end{align}
\label{eq:QLEStart}%
\end{subequations}
where $\kappa_i:=\kappa_{\text{int},i}+\kappa_{\text{ext},i}$ are the linewidths of the cavities in which $\kappa_{\text{int},i}$ and $\kappa_{\text{ext},i}$ are the intrinsic and extrinsic linewidths, respectively. Intrinsic losses and input quantum noise are associated with the uncorrelated zero-mean noise operators $\hat{a}_{\text{int},i}$ and $\hat{a}_{\text{ext},i}$, respectively, whereby we can conveniently define $\hat{a}_{\text{in},i}:=\bigl(\sqrt{\kappa_{\text{ext},i}}\hat{a}_{\text{ext},i}+\sqrt{\kappa_{\text{int},i}}\hat{a}_{\text{int},i}\bigr)/\sqrt{\kappa_i}$. The damping of the mechanical resonator is given by $\gamma_m $. The zero-mean quantum fluctuations $\hat{a}_{\text{in},i} $ and $\hat{b}_{\text{in},\text{m}}$ satisfy the correlations $\langle\hat{O}_{\text{in},i}(t)\hat{O}_{\text{in},i}^\dagger(t^\prime)\rangle=(\bar{N}_i+1)\delta(t-t^\prime)$, $\langle\hat{O}_{\text{in},i}^\dagger(t)\hat{O}_{\text{in},i}(t^\prime)\rangle=\bar{N}_i\delta(t-t^\prime)$, and $\langle\hat{O}_{\text{in},i}(t)\hat{O}_{\text{in},i}(t^\prime)\rangle=0$ where $i=1,2$ for $\hat{O}=\hat{a}$, and $i=\text{m}$ for $\hat{O}=\hat{b}$; $\bar{N}_i=1\big/\bigl\{\exp\bigl[\hbar\omega_i/(k_\text{B}T_i)\bigr]-1\bigr\}$ are the thermal photon (phonon) occupancies of the cavities (mechanical resonator) for $i=1,2$ ($i=\text{m}$) at temperature $T_i$. For $i=1$ and $2$, $\bar{N}_i$ can be obtained from a weighted sum of the mean occupation numbers associated with $\hat{a}_{\text{int},i}$ and $\hat{a}_{\text{ext},i}$.

In the strong-drive regime one can linearize these equations around the classical steady state of the cavities. We define $\delta\hat{a}_i:=\hat{a}_i-\alpha_i$ where $\alpha_i=2\mathcal{E}_ie^{\imath\phi_i}/\sqrt{4\Delta_i^2+\kappa_i^2}$ are the steady-state solutions, ignoring a small change in $\Delta_i$ due to a static shift in the position of the mechanical oscillator. These $\delta\hat{a}_i$ are the cavity field fluctuation operators, and under the linearization approximation have zero mean. When $\lvert\alpha_i\rvert\gg1$, \erefs{QLEStart} can be approximated by
\begin{subequations}
\begin{align}
\dot{\delta\hat{a}}_1&=-\bigl(\imath\Delta_1+\tfrac{\kappa_1}{2}\bigr)\delta\hat{a}_1-\imath J\delta\hat{a}_2-\imath G_1\bigl(\hat{b}+\hat{b}^\dagger)+\sqrt{\kappa_1}\hat{a}_{\text{in},1},\\
\dot{\delta\hat{a}}_2&=-\bigl(\imath\Delta_2+\tfrac{\kappa_2}{2}\bigr)\delta\hat{a}_2-\imath J\delta\hat{a}_1-\imath G_2\bigl(\hat{b}+\hat{b}^\dagger)+\sqrt{\kappa_2}\hat{a}_{\text{in},2},\ \text{and}\\
\dot{\delta\hat{b}}&=-\bigl(\imath\omega_\text{m}+\tfrac{\gamma_\text{m}}{2}\bigr)\hat{b}-\imath\bigl(G_1^\ast\delta\hat{a}_1+G_2^\ast\delta\hat{a}_2+\text{h.c.}\bigr)+\sqrt{\gamma_\text{m}}b_{\text{in},\text{m}},
\end{align}
\label{eq:QLELin}%
\end{subequations}
where $G_i=g_i\alpha_i$ are the multi-photon optomechanical coupling rates.

We choose driving frequencies such that $\Delta_i=\omega_\text{m}$. In addition, we assume operation in the sideband-resolved regime, i.e., $\omega_\text{m}\gg\kappa_i$. Under these conditions around cavity resonance frequency it is possible to use the so-called rotating-wave approximation to simplify \erefs{QLELin}. This allows to eliminate $\delta\hat{b}$ from the equations, whereby they can be approximated in the frequency domain by
\begin{equation}
-\imath\omega\begin{pmatrix}
  \delta\hat{a}_1 \\
  \delta\hat{a}_2
 \end{pmatrix}=\begin{bmatrix}
  -\imath\Delta_1-\frac{\kappa_1}{2}-\lvert G_1\rvert^2\chi_\text{m}(\omega) & -\imath J-\chi_\text{m}(\omega)G_1G_2^\ast \\
  -\imath J-\chi_\text{m}(\omega)G_1^\ast G_2&-\imath\Delta_2-\frac{\kappa_2}{2}-\lvert G_2\rvert^2\chi_\text{m}(\omega)
 \end{bmatrix}\begin{pmatrix}
  \delta\hat{a}_1 \\
  \delta\hat{a}_2
 \end{pmatrix}+\begin{pmatrix}
  \sqrt{\kappa_1}\hat{a}_{\text{in},1} \\
  \sqrt{\kappa_2}\hat{a}_{\text{in},2}
 \end{pmatrix}+\begin{pmatrix}
  -\imath G_1\sqrt{\gamma_\text{m}}\chi_\text{m}(\omega) \\
  -\imath G_2\sqrt{\gamma_\text{m}}\chi_\text{m}(\omega)
 \end{pmatrix}\hat{b}_{\text{in},\text{m}}
\end{equation}
where the mechanical susceptibility is defined as $\chi_\text{m}(\omega)=1\big/\bigl[\gamma_\text{m}/2-\imath(\omega-\omega_\text{m})\bigr]$.

\section{Conversion between cascaded and optomechanical formalisms}
In this section we give more detail on the equivalence between the two different sets of Langevin equations. Let us start by listing the two sets of expressions in frequency space. First, the optical cavities of the optomechanical system can, under the rotating-wave approximation, be described by:
\begin{equation}
-\imath\omega\begin{pmatrix}
\delta\hat{a}_1 \\
\delta\hat{a}_2
\end{pmatrix} = \begin{bmatrix}
-\imath\Delta_1-\tfrac{\kappa_1}{2}-G_1^2\chi_\text{m}(\omega) & -\imath J-G_1G_2\chi_\text{m}(\omega)e^{-\imath\phi} \\
-\imath J-G_1G_2\chi_\text{m}(\omega)e^{\imath\phi} & -\imath\Delta_2-\tfrac{\kappa_2}{2}-G_2^2\chi_\text{m}(\omega)
\end{bmatrix}\begin{pmatrix}
\delta\hat{a}_1 \\
\delta\hat{a}_2
\end{pmatrix}+\begin{pmatrix}
\sqrt{\kappa_1}\hat{a}_{\text{in},1} \\
\sqrt{\kappa_1}\hat{a}_{\text{in},2}
\end{pmatrix}+\begin{pmatrix}
-\imath G_1\sqrt{\gamma_\text{m}}\chi_\text{m}(\omega) \\
-\imath G_2\sqrt{\gamma_\text{m}}\chi_\text{m}(\omega)e^{\imath\phi}
\end{pmatrix}\hat{b}_\text{in,m},
\end{equation}
where $G_1$ is assumed to be real and non-negative and the replacement $G_2\to G_2e^{\imath\phi}$, with real and non-negative $G_2$ on the right-hand side, has already been effected. We can also perform the simple gauge transformation $\hat{b}_\text{in,m}\to\imath e^{-\imath\nu}\hat{b}_\text{in,m}$, where $\nu=\arg\{\chi_\text{m}(\Omega)\}$ and $\Omega$ is some fixed frequency of interest, to write
\begin{equation}
-\imath\omega\begin{pmatrix}
\delta\hat{a}_1 \\
\delta\hat{a}_2
\end{pmatrix} = \begin{bmatrix}
-\imath\Delta_1-\tfrac{\kappa_1}{2}-G_1^2\chi_\text{m}(\omega) & -\imath J-G_1G_2\chi_\text{m}(\omega)e^{-\imath\phi} \\
-\imath J-G_1G_2\chi_\text{m}(\omega)e^{\imath\phi} & -\imath\Delta_2-\tfrac{\kappa_2}{2}-G_2^2\chi_\text{m}(\omega)
\end{bmatrix}\begin{pmatrix}
\delta\hat{a}_1 \\
\delta\hat{a}_2
\end{pmatrix}+\begin{pmatrix}
\sqrt{\kappa_1}\hat{a}_{\text{in},1} \\
\sqrt{\kappa_1}\hat{a}_{\text{in},2}
\end{pmatrix}+\begin{pmatrix}
G_1\sqrt{\gamma_\text{m}}\tilde{\chi}_\text{m}(\omega) \\
G_2\sqrt{\gamma_\text{m}}\tilde{\chi}_\text{m}(\omega) e^{\imath\phi}
\end{pmatrix}\hat{b}_\text{in,m},
\end{equation}
defining $\tilde{\chi}_\text{m}(\omega):=\chi_\text{m}(\omega)\lvert\chi_\text{m}(\Omega)\rvert/\chi_\text{m}(\Omega)=e^{-\imath\nu}\chi_\text{m}(\omega)$.

Next, we write the cascaded system equations in frequency space. This procedure assumes that the various coefficients that appear are not time- or frequency-dependent. For generality, we shall use $\theta$ as the phase angle between $\hat{c}_1$ and $\hat{c}_2$. Thus,
\begin{equation}
-\imath\omega\begin{pmatrix}
\hat{c}_1 \\
\hat{c}_2
\end{pmatrix} = \begin{bmatrix}
-\imath\omega_1-\tfrac{\gamma_1+\kappa_1}{2} & -\imath F \\
-\imath F^\ast-\sqrt{\gamma_1\gamma_2}e^{\imath\theta} & -\imath\omega_2-\tfrac{\gamma_2+\kappa_2}{2}
\end{bmatrix}\begin{pmatrix}
\hat{c}_1 \\
\hat{c}_2
\end{pmatrix}+\begin{pmatrix}
\sqrt{\kappa_1}\hat{c}_{\text{in},1} \\
\sqrt{\kappa_2}\hat{c}_{\text{in},2}
\end{pmatrix}+\begin{pmatrix}
\sqrt{\gamma_1} \\
\sqrt{\gamma_2}e^{\imath\theta} \\
\end{pmatrix}\hat{c}_\text{in,3}.
\end{equation}
Let us compare the two sets of expressions to obtain an equivalence; we shall, for the time being, ignore the frequency-dependence of $\chi_\text{m}(\omega)$. First, it is clear that
\begin{equation}
\omega_i = \Delta_i+G_i^2\Im\{\chi_\text{m}(\omega)\}\quad(i=1,2).
\end{equation}
From the first term on the right-hand side of each expression, we can also deduce that
\begin{equation}
\gamma_i = 2G_i^2\Re\{\chi_\text{m}(\omega)\}\quad(i=1,2).
\end{equation}
Next,
\begin{equation}
F=J-\imath\chi_\text{m}(\omega)G_1G_2e^{-\imath\phi}.
\end{equation}
Continuing further, we can deduce two equations for $F^\ast$ that must hold simultaneously: $F^\ast=J-\imath G_1G_2\chi_\text{m}(\omega)e^{\imath\phi}+\imath\sqrt{\gamma_1\gamma_2}e^{\imath\theta}$ and $F^\ast=J+\imath G_1G_2\chi_\text{m}^\ast(\omega)e^{\imath\phi}$. From these two equations we deduce that
\begin{equation}
\sqrt{\gamma_1\gamma_2}e^{\imath\theta}=2G_1G_2\Re\{\chi_\text{m}(\omega)\}e^{\imath\phi}.
\end{equation}
Finally, comparing the second noise term on the right-hand side of the Langevin equations, we find
\begin{align}
\sqrt{\gamma_1} & = G_1\sqrt{\gamma_\text{m}}\lvert\chi_\text{m}(\omega)\rvert, \text{and}\\
\sqrt{\gamma_2}e^{\imath\theta} & = G_2\sqrt{\gamma_\text{m}}\lvert\chi_\text{m}(\omega)\rvert e^{\imath\phi}.
\end{align}
The equations for $\gamma_i$ ($i=1,2$) are only equivalent if
\begin{equation}
\theta=\phi,
\end{equation}
and
\begin{equation}
2\Re\{\chi_\text{m}(\omega)\}=\gamma_\text{m}\lvert\chi_\text{m}(\omega)\rvert^2.
\end{equation}
However, we note that this always holds, since
\begin{equation}
\chi_\text{m}(\omega)=\frac{1}{\tfrac{\gamma_\text{m}}{2}-\imath(\omega-\omega_\text{m})},
\end{equation}
whereby
\begin{align}
\Re\{\chi_\text{m}(\omega)\}&=\frac{\tfrac{\gamma_\text{m}}{2}}{\bigl(\tfrac{\gamma_\text{m}}{2}\bigr)^2+(\omega-\omega_\text{m})^2}, \text{and}\\
\lvert\chi_\text{m}(\omega)\rvert^2&=\frac{1}{\bigl(\tfrac{\gamma_\text{m}}{2}\bigr)^2+(\omega-\omega_\text{m})^2}.
\end{align}
Incidentally, note that $\Re\{\chi_\text{m}(\omega)\}\geq0$, which guarantees that the $\gamma_i$ ($i=1,2$) are also non-negative. It can now be seen that the two expressions are perfectly equivalent \emph{if we ignore the $\omega$-dependence of $\chi_\text{m}(\omega)$}. In order to discuss the flow of thermal noise through the system we argue as follows. We are interested in thermal state whose bandwidth $\Gamma$ is small relative to $\gamma_\text{m}$, and which is centred at frequency $\Omega$. With this in mind we can now state ($i=1,2$ throughout):
\begin{align}
\text{Cascaded system} & \leftrightarrow \text{Optomechanical platform} \nonumber\\
\hat{c}_i & \leftrightarrow \delta\hat{a}_i\\
\hat{c}_{\text{in},i} & \leftrightarrow \hat{a}_{\text{in},i}\\
\hat{c}_{\text{in},3} & \leftrightarrow \hat{b}_\text{in,m}\\
\omega_i & \leftrightarrow \Delta_i+G_i^2\Im\{\chi_\text{m}(\Omega)\}=\Delta_i+\frac{G_i^2(\Omega-\omega_\text{m})}{\bigl(\tfrac{\gamma_\text{m}}{2}\bigr)^2+(\Omega-\omega_\text{m})^2} \\
\gamma_i & \leftrightarrow 2G_i^2\Re\{\chi_\text{m}(\Omega)\}=\frac{G_i^2\gamma_\text{m}}{\bigl(\tfrac{\gamma_\text{m}}{2}\bigr)^2+(\Omega-\omega_\text{m})^2} \\
\theta & \leftrightarrow \phi \\
F & \leftrightarrow J-\imath\chi_\text{m}(\Omega)G_1G_2e^{-\imath\phi}=J-\frac{\imath G_1G_2e^{-\imath\phi}}{\tfrac{\gamma_\text{m}}{2}-\imath(\Omega-\omega_\text{m})}
\end{align}
At this stage, note that none of these coefficients depends on $\omega$, as formally required for the expressions derived using the cascaded systems formalism to be valid. It is easy to read off that perfect non-reciprocity requires $F=0$, i.e.,
\begin{equation}
J=\frac{G_1G_2}{\sqrt{\bigl(\tfrac{\gamma_\text{m}}{2}\bigr)^2+(\Omega-\omega_\text{m})^2}},
\end{equation}
with $\phi$ chosen appropriately, constrained by the demand that $J$ is real.

In the large-bandwidth limit ($\gamma_\text{m}\to\infty$) we obtain a perfect equivalence, since $\chi_\text{m}(\omega)=2/\gamma_\text{m}$ is then no longer a function of frequency. Under these conditions, we can write
\begin{align}
\text{Cascaded system} & \leftrightarrow \text{Optomechanical platform} \nonumber\\
\hat{c}_i & \leftrightarrow \delta\hat{a}_i\\
\hat{c}_{\text{in},i} & \leftrightarrow \hat{a}_{\text{in},i}\\
\hat{c}_{\text{in},3} & \leftrightarrow \hat{b}_\text{in,m}\\
\omega_i & \leftrightarrow \Delta_i \\
\gamma_i & \leftrightarrow \frac{4G_i^2}{\gamma_\text{m}} \\
\theta & \leftrightarrow \phi \\
F & \leftrightarrow J-\frac{2\imath G_1G_2e^{-\imath\phi}}{\gamma_\text{m}}
\end{align}

\section{Occupation numbers}
Based on the assumption that none of the coefficients entering the cascaded system calculation is time- or frequency-dependent, it is relatively straightforward to obtain the steady-state occupation numbers for the two oscillators. First, define $\Delta:=\omega_2-\omega_1$ for simplicity, and let $\bar{N}_1$, $\bar{N}_2$, and $\bar{N}_3$ be the occupation numbers for the baths defined by $\hat{b}_{\text{in},1}$, $\hat{b}_{\text{in},2}$, and $\hat{b}_\text{in,m}$, respectively. Furthermore, let us simplify matters by taking $\kappa_1=\kappa_2=\gamma_1=\gamma_2=:\kappa$ Then,
\begin{equation}
\bar{n}_1=\frac{2\lvert F\rvert^2(\bar{N}_1+\bar{N}_2+\bar{N}_3)+\Re\{Fe^{\imath\theta}\}\Delta(\bar{N}_1-\bar{N}_3)+2\Im\{Fe^{\imath\theta}\}^2\bar{N}_3+2\Im\{Fe^{\imath\theta}\}\kappa(\bar{N}_1+3\bar{N}_3)+(4\kappa^2+\Delta^2)(\bar{N}_1+\bar{N}_3)}{2[3\lvert F\rvert^2+4\kappa(\kappa+\Im\{Fe^{\imath\theta}\})+\Im\{Fe^{\imath\theta}\}^2+\Delta^2]},
\end{equation}
and
\begin{multline}
\bar{n}_2=\frac{2\lvert F\rvert^2(\bar{N}_1+\bar{N}_2+\bar{N}_3)-\Re\{Fe^{\imath\theta}\}\Delta(\bar{N}_2-\bar{N}_3)+2\Im\{Fe^{\imath\theta}\}^2\bar{N}_3+2\Im\{Fe^{\imath\theta}\}\kappa(\bar{N}_2+3\bar{N}_3)+(4\kappa^2+\Delta^2)(\bar{N}_2+\bar{N}_3)}{2[3\lvert F\rvert^2+4\kappa(\kappa+\Im\{Fe^{\imath\theta}\})+\Im\{Fe^{\imath\theta}\}^2+\Delta^2]}\\
+\frac{\kappa(2\Im\{Fe^{\imath\theta}\}+\kappa)(\bar{N}_1-\bar{N}_3)}{3\lvert F\rvert^2+4\kappa(\kappa+\Im\{Fe^{\imath\theta}\})+\Im\{Fe^{\imath\theta}\}^2+\Delta^2}.
\end{multline}
For perfect non-reciprocity we set $F=0$ and obtain
\begin{equation}
\bar{n}_1=\tfrac{1}{2}(\bar{N}_1+\bar{N}_3),
\end{equation}
and
\begin{equation}
\bar{n}_2=\tfrac{1}{2}(\bar{N}_2+\bar{N}_3)+\frac{\kappa^2(\bar{N}_1-\bar{N}_3)}{4\kappa^2+\Delta^2}.
\end{equation}
Simplifying the latter further for the resonant case, $\Delta=0$, we obtain
\begin{equation}
\bar{n}_2^{(\Delta=0)}=\tfrac{1}{4}(\bar{N}_1+2\bar{N}_2+\bar{N}_3)=\tfrac{1}{2}(\bar{N}_2+\bar{n}_1).
\end{equation}
The seemingly anomalous factor $\tfrac{1}{2}$ in the expressions for $\bar{n}_1$ and $\bar{n}_2^{(\Delta=0)}$ comes from the fact that the two oscillators are both connected to a third bath. Indeed, for a fair comparison, we can consider the two oscillators connected to two baths each, but devoid of any direct coupling or common baths. In this (``disconnected'') scenario, which is physically equivalent to taking $\lvert\Delta\rvert\to\infty$ in the above expressions whilst keeping $F$, $\kappa$, and $\bar{N}_i$ ($i=1,2,3$) fixed, the steady-state occupation numbers are, instead,
\begin{align}
\bar{m}_1&=\tfrac{1}{2}(\bar{N}_1+\bar{N}_3), \text{and}\\
\bar{m}_2&=\tfrac{1}{2}(\bar{N}_2+\bar{N}_3).
\end{align}
Thus, if we again allow $\Delta$ to be general,
\begin{align}
\bar{n}_1&=\bar{m}_1, \text{and}\\
\bar{n}_2&=\bar{m}_2+\frac{\kappa^2(\bar{N}_1-\bar{N}_3)}{4\kappa^2+\Delta^2}.
\end{align}
This very clearly shows that, whatever the value of $\bar{N}_1-\bar{N}_2$, we find an \emph{increase} (\emph{decrease}) in $\bar{n}_2$ over the disconnected case for $\bar{N}_1>\bar{N}_3$ ($\bar{N}_1<\bar{N}_3$), whereas $\bar{n}_1$ is unaffected by the presence of the other oscillator. It is interesting to note that this conclusion remains unchanged if we have $\bar{N}_2=\bar{N}_1$.

For general $F$ and $\Delta$ we find
\begin{align}
\bar{n}_1-\bar{m}_1&=\frac{\lvert F\rvert^2(-\bar{N}_1+2\bar{N}_2-\bar{N}_3)-\bigl[\Im\{Fe^{\imath\theta}\}(2\Im\{Fe^{\imath\theta}\}+\kappa)-\Re\{Fe^{\imath\theta}\}\Delta\bigr](\bar{N}_1-\bar{N}_3)}{2[3\lvert F\rvert^2+4\kappa(\kappa+\Im\{Fe^{\imath\theta}\})+\Im\{Fe^{\imath\theta}\}^2+\Delta^2]}, \text{and}\\
\bar{n}_2-\bar{m}_2&=\frac{\lvert F\rvert^2(2\bar{N}_1-\bar{N}_2-\bar{N}_3)-\bigl[\Im\{Fe^{\imath\theta}\}(2\Im\{Fe^{\imath\theta}\}+\kappa)+\Re\{Fe^{\imath\theta}\}\Delta\bigr](\bar{N}_2-\bar{N}_3)+2\kappa(2\Im\{Fe^{\imath\theta}\}+\kappa)(\bar{N}_1-\bar{N}_3)}{2[3\lvert F\rvert^2+4\kappa(\kappa+\Im\{Fe^{\imath\theta}\})+\Im\{Fe^{\imath\theta}\}^2+\Delta^2]}.
\end{align}

\begin{figure}[t]
 \includegraphics[width=0.35\linewidth]{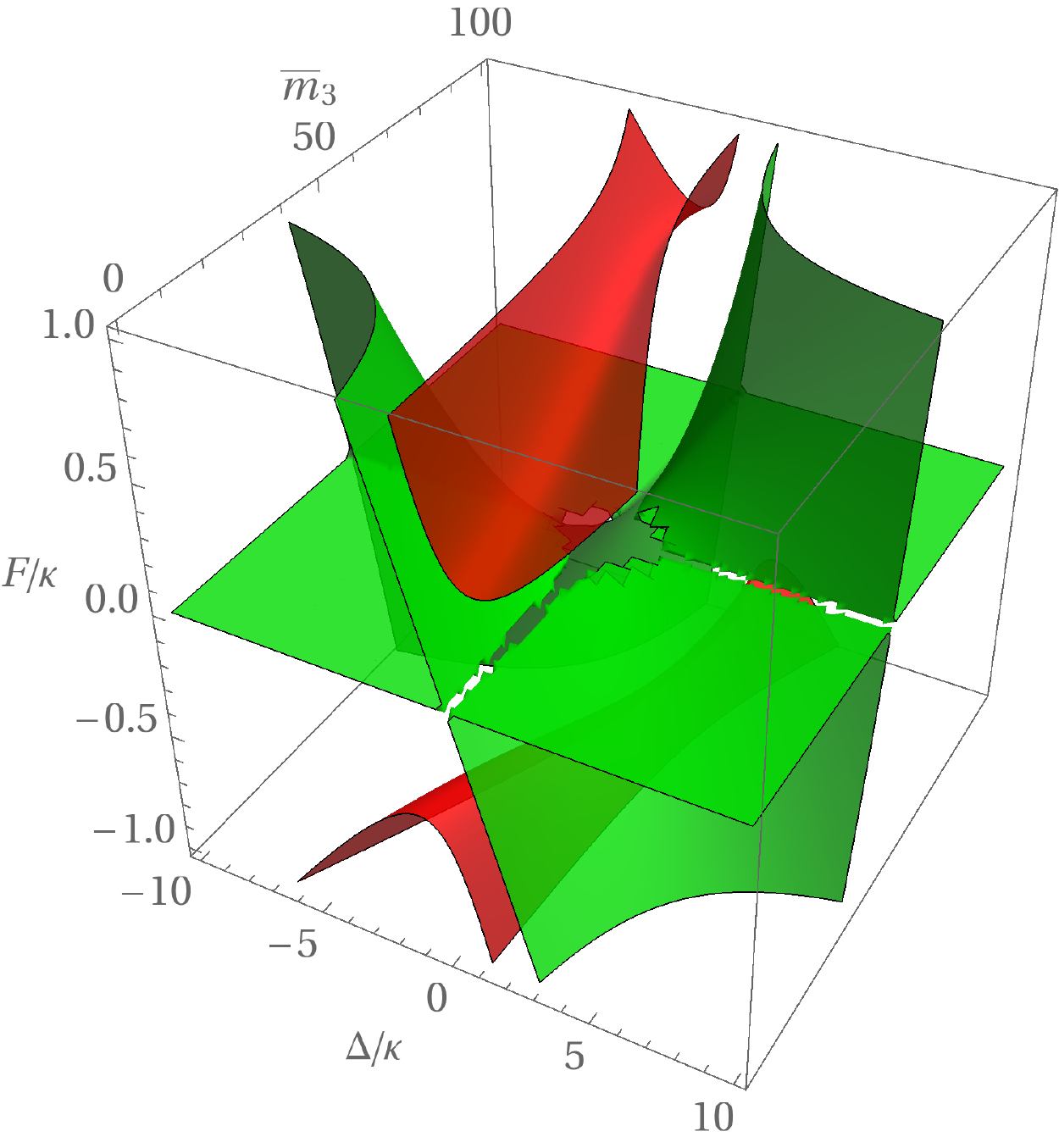}
 \includegraphics[width=0.35\linewidth]{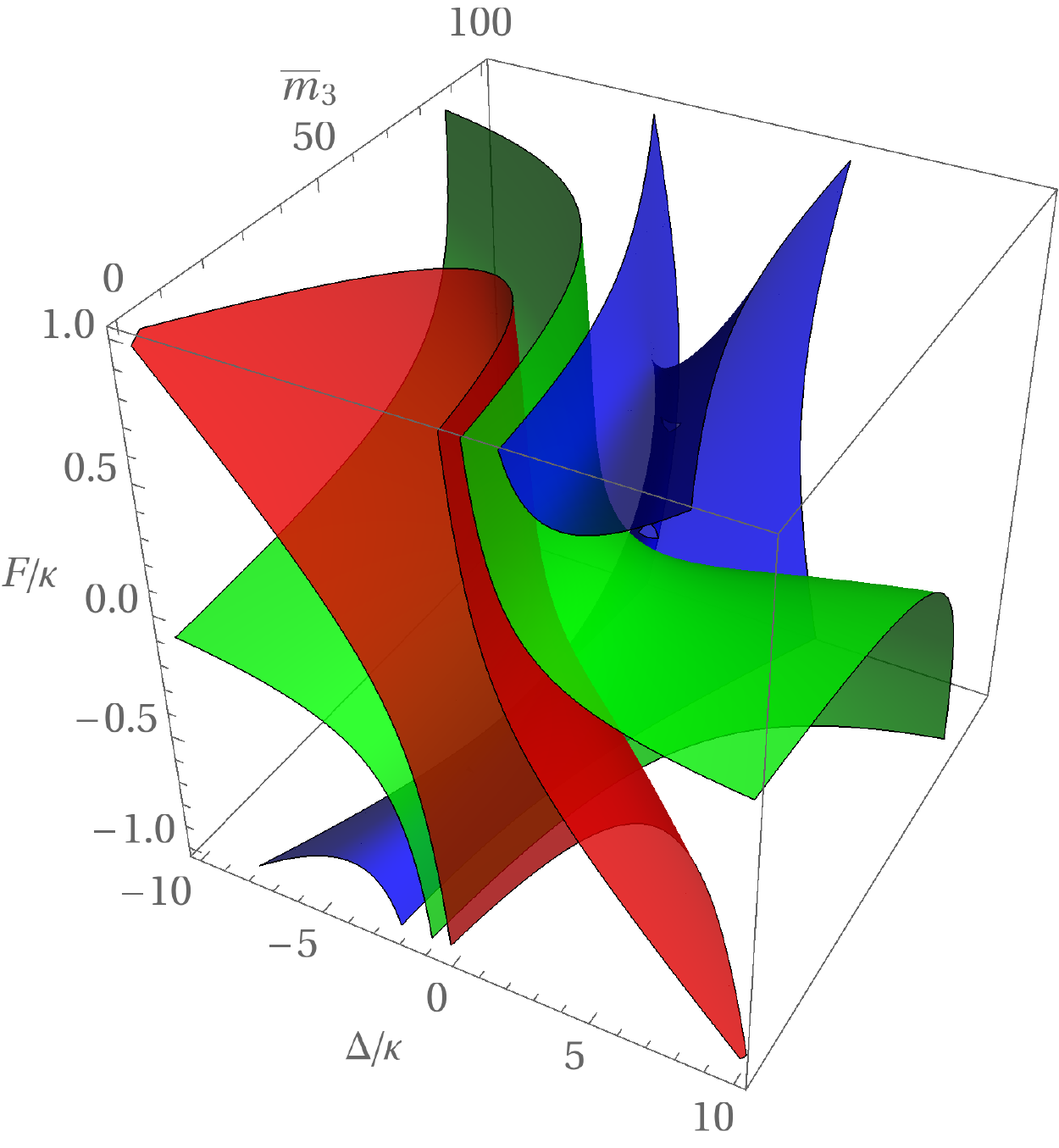}
 \caption{(Color online) Change in occupation number of the first oscillator ($\Delta n_1$, left) and second oscillator ($\Delta n_2$, right), as a function of the detuning $\Delta$ between the two oscillators, the breaking of non-reciprocity $F$, and the occupation number $\bar{m}_3$ of the common bath. The red, green, and blue surfaces are, respectively, where $\Delta n_i=10$, $0$, and $-10$ ($i=1,2$). Fig.~2 in the main text represents a slice of the right-hand plot for $F=0$. Note that $\Delta n_1=0$ when $F=0$. ($\phi=0$, $\bar{m}_1=50$, $\bar{m}_2=100$.)}
 \label{fig:Dn}
\end{figure}
We can generalize Fig.~2 in the main text for the case of imperfect non-reciprocity, obtaining \fref{Dn}. These two figures show the versatility of the system at hand, where the control parameters $\Delta$, $F$, and $\bar{m}_3$ can be used to set the temperature difference of either oscillator with respect to the disconnected system. As expected, the first oscillator can never be cooled, but it is indeed possible to cool the second oscillator. This shows that reduced net thermal noise flow can be set up to oscillator 2, despite the fact that, when $F=0$, all coherent signals flow from oscillator 1 to oscillator 2.

\section{Rate of flow of excitations into and out of the baths}
In this section we will briefly summarize a technique that can be used to obtain knowledge of the full counting statistics of the exchange of excitations between a quantum system and a heat bath. The development of this technique can be traced in recent literature (see Refs.\ \cite{2015PhRvA..92a3844P, 2016NJPh...18a3009P, 2016JSMTE..06.3203P} and references therein); the focus here is on its application to Gaussian states evolving under the action of dynamics that preserves their Gaussian nature (see Appendix B in Ref.\ \cite{2016JSMTE..06.3203P}). The basis of the technique rests on the definition of a biased covariance matrix $V_s$, which under steady-state conditions satisfies the relation
\begin{equation}
0=[A-F_-(s)]\cdot V_s+V_s\cdot[A-F_-(s)]^\text{T}+V_s\cdot F_+(s)\cdot V_s+N,
\label{eq:RiccatiVs}
\end{equation}
where $A$ is the drift matrix and $N$ the noise matrix that is obtained from the noise terms entering the corresponding Langevin equations; both are defined in the main text. We define the auxiliary matrices $F_\pm(s)$ through the relation
\begin{figure}[t]
 \includegraphics[width=\linewidth]{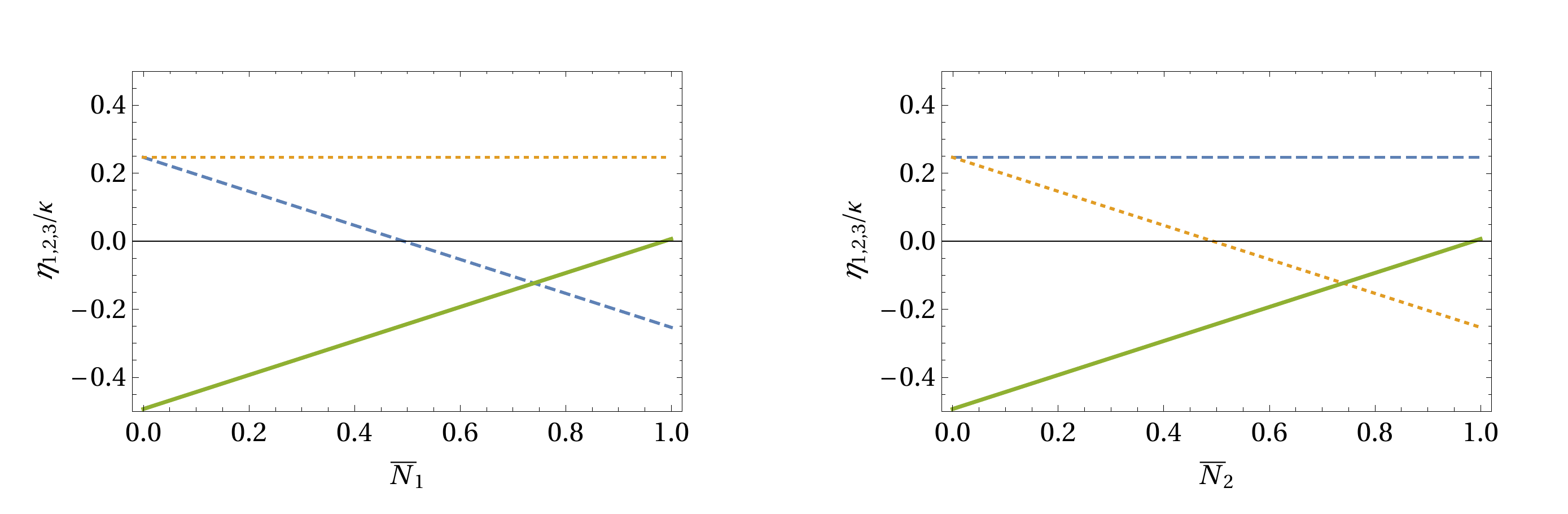}
 \caption{Rate of flow of excitations between the system and its three baths. In both figures, the blue (dashed) curves correspond to $\eta_1$, the orange (dotted) curves to $\eta_2$, and the green (solid) curves to $\eta_3$. For convenience of presentation the flows are normalized by dividing by $\kappa$. The left plot shows the situation for $\bar{N}_2\approx0$, and the right plot for $\bar{N}_1\approx0$. In both cases $\bar{N}_3\approx0.5$ and the other parameters are as in the main text.}
 \label{fig:ExcitationFlows}
\end{figure}
\begin{equation}
F_\pm(s)=\bigoplus_{j=1}^N\delta_{i,j}\begin{bmatrix}
f_{j\pm}(s) & 0 \\
0 & f_{j\pm}(s)
\end{bmatrix},
\end{equation}
where $i=1,2,3$ is the noise channel of interest and $f_{j\pm}(s)=\gamma_j\bigl[(\bar{N}_j+1)(e^{-s}-1)\pm\bar{N}_j(e^s-1)\bigr]$, where $\gamma_j$ is the rate through which the system is coupled to bath $j$ and $\bar{N}_j$ the mean number of excitations of this bath. The ordinary case, where $V_s$ reduces to the usual covariance matrix $V$, results from taking $s=0$, whereupon $f_{j\pm}(0)=0$, $F_\pm(0)=0$, and the algebraic Riccati equation, \eref{RiccatiVs}, reduces to the usual steady-state Lyapunov equation, as used the main text. The full counting statistics of the counting process associated with the excitations being exchanged between the system and bath $i$ can be obtained through the large-deviation function
\begin{equation}
\theta(s)=\tfrac{1}{2}\Tr\{F_+(s)\cdot V_s-F_-(s)\},
\end{equation}
with the $n$\textsuperscript{th} derivative of $\theta(s)$ evaluated at $s=0$ being related to the $n$\textsuperscript{th} moment of the counting process, $\eta^{(n)}$, through the relation
\begin{equation}
\eta^{(n)}=(-1)^n\bigl[\partial_s^n\theta(s)\rvert_{s=0}\bigr].
\end{equation}
For convenience we drop the superscript when referring to the first moment and define $\eta_i$ ($i=1,2,3$) to be the first moment of the counting process---i.e., the average rate of flow of excitations---between the system and bath $i$. A concise expression can be obtained for these first moments that does not make reference to $V_s$ directly:
\begin{equation}
\eta_i=-\tfrac{1}{2}\Tr\{F_+^\prime\cdot V-F_-^\prime\},
\end{equation}
where $F_\pm^\prime$ is the first derivative of $F_\pm$, evaluated at $s=0$:
\begin{equation}
F_\pm^\prime=\bigoplus_{j=1}^N\delta_{i,j}\begin{bmatrix}
f_{j\pm}^\prime & 0 \\
0 & f_{j\pm}^\prime
\end{bmatrix},
\end{equation}
with $f_{j\pm}^\prime=-\gamma_j\bigl[\bar{N}_j(1\mp1)+1\bigr]$. Applying this procedure to the generic system described in the main text, we obtain rather unwieldy expressions. However, in the simplified situation where the $\gamma_i$ and $\kappa_i$ are all equal to $\kappa$, and $F=0$, we find
\begin{subequations}
\begin{align}
\eta_1&=\kappa(\bar{n}_3-\bar{n}_1),\\
\eta_2&=\kappa\biggl[\frac{2\kappa^2}{4\kappa^2+\Delta^2}(\bar{n}_1-\bar{n}_3)+(\bar{n}_3-\bar{n}_2)\biggr],\ \text{and}\\
\eta_3&=\kappa\biggl[\frac{2\kappa^2}{4\kappa^2+\Delta^2}(\bar{n}_3-\bar{n}_1)+(\bar{n}_1-\bar{n}_3)+(\bar{n}_2-\bar{n}_3)\biggr],
\end{align}
\end{subequations}
such that $\sum_j\eta_j=0$, as required when accounting for all the heat baths connected to a system. Under these simplified conditions we can see that $\eta_1$ ($\eta_2$) does \emph{not} depend on $\bar{N}_2$ ($\bar{N}_1$). This is shown explicitly in \fref{ExcitationFlows} for the parameters used in the main text. Thus, (i)~any excess excitations flowing out of the common heat bath into system $1$ is cancelled by an equal flow of excitations between systems $1$ and $2$; and (ii)~despite the fact that $\bar{n}_2$ increases as $\bar{N}_1$ is increased, this is not reflected explicitly in the net flow of excitations between system $2$ and its own heat bath.

\twocolumngrid

\end{document}